\documentclass[
reprint,
superscriptaddress,
amsmath,amssymb,
aps,
prb,
]{revtex4-2}

\usepackage{amsmath}
\usepackage[utf8]{inputenc}
\usepackage{siunitx} 
\usepackage{xcolor}
\usepackage{natbib}
\usepackage{graphicx}
\usepackage{braket}
\usepackage{comment}
\usepackage{bm}
\usepackage{epstopdf}

\usepackage{graphicx}
\usepackage{dcolumn}
\usepackage{bm}
\usepackage{comment}

\usepackage{braket}
\usepackage{color}
\usepackage{array}
\newcolumntype{?}{!{\vrule width 1pt}}

\begin{document}



\title{Combined tight-binding and configuration interaction study of unfolded electronic structure of G-color center in Si}

\date{\today}
\author{Jakub Valdhans}%
    \email{valdhans@physics.muni.cz}
    \affiliation{Department of Condensed Matter Physics, Faculty of Science, Masaryk University, Kotl\'a\v{r}sk\'a~267/2, 61137~Brno, Czech~Republic}

\author{Petr Klenovsk\'{y}}%
    \email{klenovsky@physics.muni.cz}
    \affiliation{Department of Condensed Matter Physics, Faculty of Science, Masaryk University, Kotl\'a\v{r}sk\'a~267/2, 61137~Brno, Czech~Republic}
    \affiliation{Czech Metrology Institute, Okru\v{z}n\'i 31, 63800~Brno, Czech~Republic}

\begin{abstract} 
We have theoretically studied the G-center in bulk silicon material using the empirical tight-binding model for calculations of unfolded band structures with configuration interaction correction for the exciton at $\Gamma$ point of the Brillouin zone. The G-center in B configuration (emissive) being a candidate structure as the telecom single- and entangled-photon source has two substitutional carbons and one interstitial atom embedded into the bulk in six equally possible configurations. Taking the advantage of the low computation effort of the tight-binding and unfolding approach, it is possible to calculate and analyze the behavior of a variety of the electronic configurations. Our tight-binding model is able to describe not only the behavior of the G-center in the silicon bulk but using the unfolding approach it can also pinpoint the contributions of different elements of the supercell on the final pseudo-band structure. Moreover, the configuration interaction correction with single-particle basis states computed by our unfolded tight-binding model predicts a very small fine-structure splitting of the ground state exciton both for bright and dark doublet in the studied system. That underscores the possibility of the silicon G-center to become a very good emitter of single and entangled photons for quantum communication and computation applications.

\end{abstract}

\maketitle

\section{Introduction}
\label{sec:intro}
\subsubsection{\textcolor{black}{G-color center in the emissive B-type configuration}}
One of the most important semiconductors is silicon and its impact has played a key role in the electronics and photovoltaics industries.~\cite{ruzova_kniha,Gc_Beaufils2018_navic,Gc_B_cof_Shainline2007} However its indirect band gap makes silicon an inefficient light emitter. To improve that, implanting silicon with interstitial atoms, called color-centers, has been identified as a promising route. For one type of color-centers in Si, the G-center in the emissive B-type configuration, an emission of the zero-phonon line (ZPL) at 1.278 - 1.280 µm (0.970 - 0.969 eV)~\cite{Gc_OSC_2type_II,Gc_Durand2024_main,Gc_B_cof_Shainline2007,Gc_Beaufils2018_navic,Density_of_Gc2011,Gc_Durand2024_OSC,Gc_energ_Davies1983,Gc_energ_Thonke1981} corresponding to the telecom O-band~\cite{o_symetrii_Gc} has been measured at low-temperatures. That makes the G-color centers in Si a very promising candidate as a potential quantum light emitter.~\cite{Fox2024,Hollenbach2020,Rahmouni2024,Liu2024,Day2024}
\subsubsection{\textcolor{black}{Empirical tight-binding model}}
\indent The electronic band structure of silicon is changed from the indirect band gap to the direct band gap by the presence of the G-center embedded in silicon bulk. The empirical tight-binding (ETB) model~\cite{Harrison1973,Vogl1983,Jancu1998} is used here to simulate this behavior. The ETB model is a variational one particle method using empirical bulk parameters (ETB parameters) replacing terms in the Hamiltonian by plain numbers, set before calculation \{these are usually results of fits to density functional theory (DFT) calculations of bulk band structure\} in bases of atomic orbitals localized on atomic sites in the structure. We divide the ETB parameters into on-site (localized on the main diagonal of the Hamiltonian), which correspond to the energy of an electron being on a certain atomic orbital and off-site (localized off the main diagonal of the Hamiltonian) corresponding to the hopping energy amplitude of an electron between atomic orbitals localized on different atomic sites. In order to describe the indirect band gap in bulk silicon, the extended sp$^3$d$^5$s* base (ten orbitals) is used in the calculations.~\cite{Jancu1998} An advantage of the ETB model is the improved numerical feasibility of calculations (note,~e.g., that a calculation for a large structure consisting of hundreds of atoms can take weeks to complete on the supercomputer in case of DFT calculations, while results for a similar structure could be obtained in tens of minutes on a regular laptop in case of ETB calculations). \textcolor{black}{Another advantage is that the ETB model makes possible a direct use of the wave functions physically composed from localized atomic orbitals unlike to DFT where the wave functions are used as an instrument to optimize the electron density~\cite{DFT_theory2002}. In this article, the wave functions from the ETB model were used for the unfolding approach to extract more information about the calculated electronic structure. Moreover, the ETB model possibly allows calculations of the electronic structure of hybrid systems composed of,~e.g., color-center and quantum dots~\cite{Lubotzky2024,Katsumi2025}, which are currently not accessible using DFT.}
\subsubsection{\textcolor{black}{Unfolding approach}}
\indent  The carbon G-center is embedded in the silicon supercell (SC) leading to a folded band structure when ETB calculations are employed without any further improvements. To analyze the folded band structure, the band unfolding approach~\cite{Boykin2016} is used in this work to distinguish the G-center and bulk-like behavior. The unfolding approach is based on the transformation into new basis of an inverse Fourier transformation (IFT) of primitive cell (PC) eigenstates. From the normalization condition of that transformation, we can introduce a new parameter, the weight representing the probabilities of the bulk-like behavior of states in the unfolded band structure. However, the unfolding approach expects a fully periodic SC structure (like for the bulk semiconductor) which is unfortunately not satisfied in the case of the G-center. Hence, in this work we introduce a method to separate contributions of periodic behavior of bulk and aperiodic behavior caused by the G-center.
\subsubsection{\textcolor{black}{Multi-particle Coulomb interaction correction, the configuration interaction}}
\indent  To incorporate the electron-electron interaction into one particle ETB model, the configuration interaction (CI)~\cite{Bryant1987,Schliwa:09,Klenovsky2017} is employed by introducing Slater determinants for the single exciton. That finally leads to the calculations of the fine-structure splitting (FSS) for dark and bright exciton in silicon G-centers.
\subsubsection{\textcolor{black}{Article contents}}
\indent  We divided the theoretical part in section II. into four parts: In \textcolor{black}{II.A}, we introduce detailed properties of the G-center and its atomic configuration. In \textcolor{black}{II.B}, we build the Hamiltonian for SC with the G-center using ETB model for the calculation of the electronic structure. In \textcolor{black}{II.C}, we implement unfolding approach for folded electronic structure to distinguish periodic and aperiodic behavior of SC and in \textcolor{black}{II.D}, we incorporate electron-electron Coulomb interaction using CI. Results of calculations are described in section III. followed by discussion and conclusions in section IV.
\section{Structural properties and Theory model}
\label{sec:outline}
\subsection{\textcolor{black}{G-center ($\text{C}_{\mathrm{Si}}{-}\text{Si}_{\mathrm{i}}{-}\text{C}_{\mathrm{Si}} \text{ complex}$) in Si bulk}}
%
\begin{figure*}[htbp]
    \includegraphics[width=160mm]{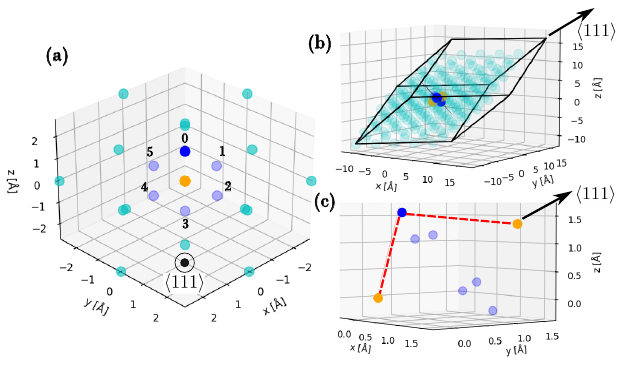}	
	\caption{The structure of the supercell (SC) with the G-center. (a) Six positions of the silicon interstitial atom Si$_\textrm{i}$ (dark blue and violet) with $i$=0-5 and two carbon substitutional atoms C (orange) aligned along the $\langle 111 \rangle$ crystal direction in the Si bulk (light blue). (b)~The SC with the G-center and depicted periodic boundary conditions by the black parallelogram. The structure of $5\times 5\times5$ primitive cells (PCs) contains 251 atoms (248 Si, 2 C$_\textrm{sub}$, 1 Si$_\textrm{int}$). 
(c) Two bonds to nearest neighbors (NNs) from Si$_\textrm{i=0}$ (blue) to C (orange) are depicted by red dashed lines.}
	\label{fig:Struct}
\end{figure*}
%
The G-center (in B-type emissive configuration~\cite{Gc_B_cof_Shainline2007,Gc_bond_length2014}) is composed of two substitutional carbon atoms with bonds pointing in the $\langle 111 \rangle$ crystal direction and one interstitial silicon atom.~\cite{Gc_Durand2024_main,Gc_B_cof_Shainline2007} The interstitial atom is located in the plane in the middle between the substitutional atoms being at one of six possible sites, as shown in Fig.~\ref{fig:Struct}~(a). Each possible interstitial atom site has only two covalent bonds~\cite{o_symetrii_Gc,Gc_Durand2024_main} to its nearest neighbor (NN) carbon atoms, depicted by red dashed lines in Fig.~\ref{fig:Struct}~(c). The G-center resides in the middle of the structure. The surrounding bulk silicon atoms and two carbon atoms have four NNs each. Periodic boundary conditions are used, represented by the black parallelogram in Fig.~\ref{fig:Struct}~(b). The frozen defect has $\textrm{C}_\textrm{1h}$ symmetry and in an unperturbed structure with the lattice constant of crystalline silicon ${a}_{\textrm{Si}}= 5.431$~\AA, the G-center remains in one of six possible equivalent sites, see Fig.~\ref{fig:Struct}(a).~\cite{o_symetrii_Gc,Gc_Durand2024_main} We use ETB model for calculating the band structure of supercell containing the G-center in the middle of the structure. 
\subsection{Empirical tight-binding model}
\begin{figure}[htbp]
	\includegraphics[width=85mm]{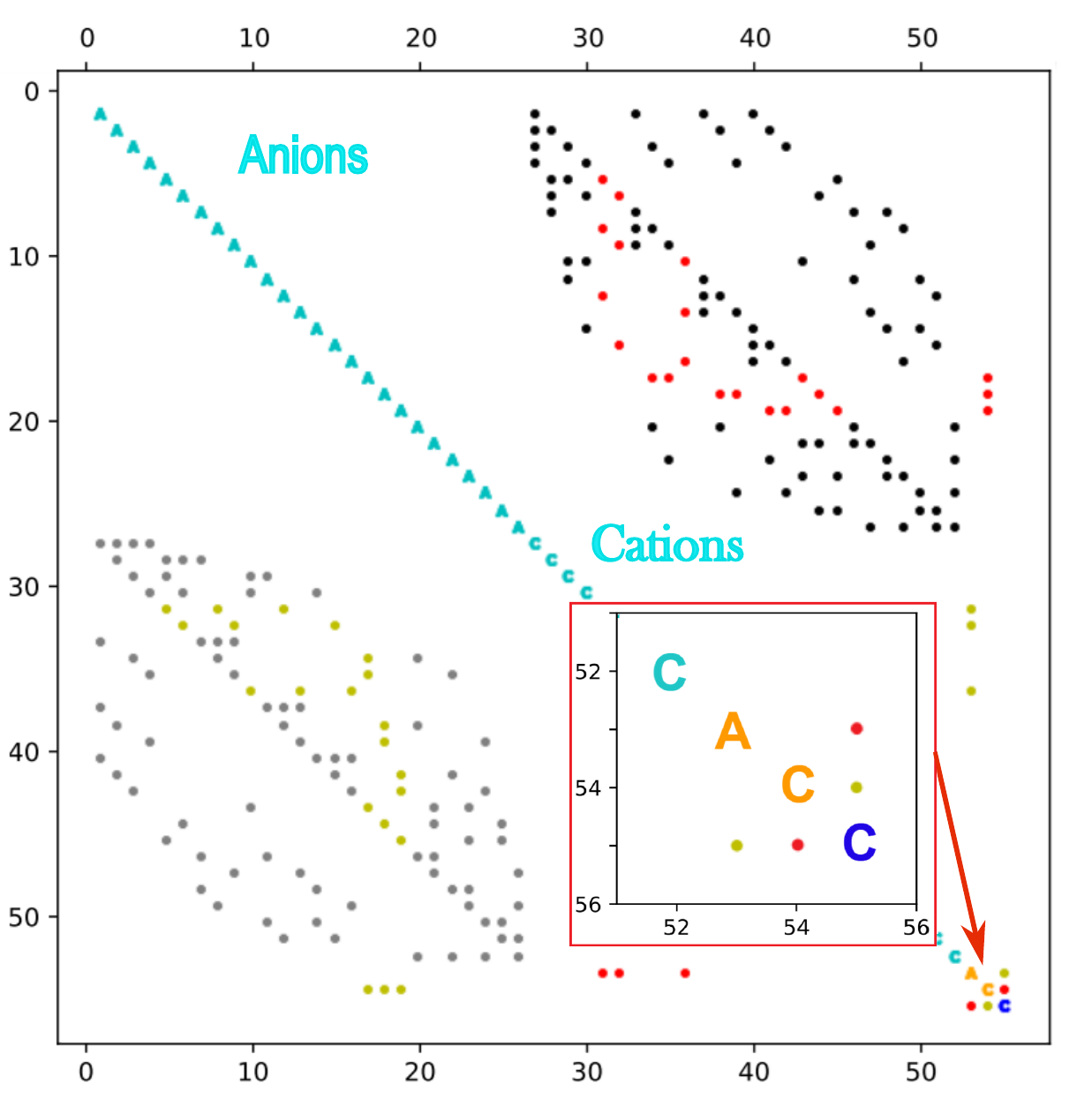}
	\caption{The Hamiltonian for the supercell with the G-center containing 55 atoms (52 Si, 2 C$_\textrm{sub}$, 1 Si$_\textrm{int}$). Red (AC) and yellow (CA) 20x20 off-site ETB matrices represent those where VCA was used and black (AC) and gray (CA) where VCA was not employed. Note that some on-site ETB matrices are also modified (not shown in this figure) due to VCA (those which have at least one NN changed by VCA). On the main diagonal, there are 20x20 on-site ETB matrices for anions (A) and cations (C) where different colors represent different types of atoms; light-blue (Si bulk), orange (C$_\textrm{sub}$) and dark-blue (Si$_\textrm{int}$).}
	\label{fig:Ham}
\end{figure}

The ETB model with sp$^3$d$^5$s* basis and spin-orbit (SO) interaction is used in calculations of the electronic band structure of the G-center. Besides one particle approximation in this model, we use the following approximations: (i)~empirical parameters with SO interaction for Si and C for low temperatures (around 0~K) taken from Ref.~\cite{Jancu1998}, (ii)~the NN coupling only, and two-center bond model,~\cite{Slater_a_koster1928,dipl_vald,ruzova_kniha} (iii)~periodic boundary conditions,~\cite{boundary_condition2004,dipl_vald} (iv)~strain effects considered using the Harrison-rule,~\cite{harrison1989electronic,Jancu1998,Harrison_rule1979} and (v)~virtual-crystal approximation (VCA) and considering the band offset.~\cite{dipl_vald} For more details on the implementation of ETB model (proper definition) and details about used approximations above, see Ref.~\cite{dipl_vald}. 
\\
\indent 
We build the Hamiltonian as in Fig.~\ref{fig:Ham} from $20\times 20$ on-site (on the main diagonal, $H_{{A}}$ and $H_{{C}}$) and off-site (off the main diagonal, $H_{{A}{C}}$ (black and red) and $H_{{C}{A}}$ (gray and yellow)) ETB matrices. For definitions of matrices, see Refs.~\cite{dipl_vald,supTeo}.
The interstitial cation atom (blue) has two covalent bonds to the carbon anion and the carbon cation (both orange), see inset of Fig.~\ref{fig:Ham}. Notice that in the inset, there is a bond between cation C$_\textrm{sub}$ (orange) and cation Si$_\textrm{int}$ (blue). Since ETB parameters are considered only between cation and anion and not between two cations, we changed the bond in the Hamiltonian as cation C$_\textrm{sub}$ (orange) a and anion Si$_\textrm{int}$ (blue), see inset of Fig.~\ref{fig:Ham}.
%
\begin{figure}[htbp]
	\includegraphics[width=70mm]{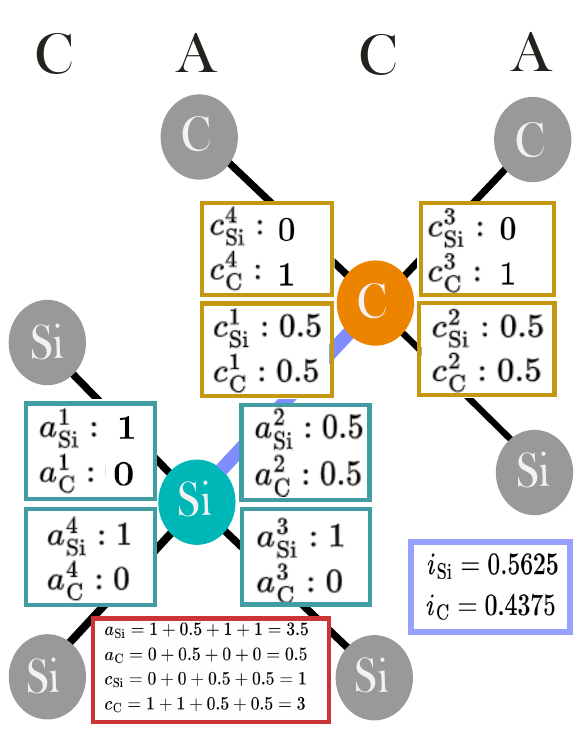}
	\caption{The determination of the weight $i_{\alpha}$ of the bond (Eq.~\eqref{VCA_off}) for an example between silicon anion and carbon cation with their NNs. Notice that final bond is composed of two off-site ETB parameters with weights $i_\textrm{Si}=0.5625$ and $i_\textrm{C}=0.4375$.}
	\label{fig:VCA_det}
\end{figure}
%
%
\subsubsection{Approximations for off-site ETB matrices}
VCA is used in the case when atoms (e.g. Si atoms) are influenced by their NNs composed of different kind of atoms (e.g. when NNs are C atoms). In that case, an average of the ETB parameters for different atom species is used to take that into account. 
\\
\indent Namely, VCA is considered for off-site matrices as
\begin{equation}\label{VCA_off}
\begin{split}  
&H_{{A}{C}}=i_{\alpha}\cdot H_{{A}{C}}^\alpha,\\
&H_{{C}{A}}=i_{\alpha}\cdot H_{{C}{A}}^\alpha,\\
&i_{\alpha}=\frac{1}{2}\left(\frac{a_\alpha}{\sum_\alpha a_\alpha}+\frac{c_\alpha}{\sum_\alpha c_\alpha}\right),
\end{split}
\end{equation}
 where $i_{\alpha}$ represents the weights for one NN around an anion (AC bond) or a cation (CA bond), with $\alpha$ enumerating different types of off-site parameters, $\alpha\in\{\textup{Si-Si},\textup{C-C}\}$; Einstein sum notation is used in the first two equations in~\eqref{VCA_off}. The weight $a_\alpha/\sum_\alpha a_\alpha$ ($c_\alpha/\sum_\alpha c_\alpha$) has the meaning of determining the type of an anion (cation) according to the influence of its NNs. See also Fig.~\ref{fig:VCA_det}.
 \\
\indent  To determine the factor $a_\alpha$, we go through all NNs of the anion. In one case, we increase the $\alpha$-type of $a_\alpha$ by 1 if the anion with the $\alpha$-type and its NN with the $\alpha'$-type are from the same structure described by $\alpha$-type ETB parameters, $\alpha=\alpha'$ \{e.g. a case of Si (anion) and Si (its NN cation) described by Si-Si ETB parameters\}. In the other case, we increase the $\alpha$-type and the $\alpha'$-type of $a_\alpha$ by 0.5 if the anion and its NN are not from the same structure, $\alpha\neq\alpha'$ \{e.g. a case of Si (anion) and C (its NN cation) where Si-C parameters are not available\}. With $a_\alpha$ marking the process, it follows that
%
%
\begin{equation}\label{VCA_off_param}
\textcolor{black}{a_\alpha = \sum_{n=1}^{n_{\rm NNs}} \left[ \delta_{\alpha, \alpha'_n} + 0.5 \left(1 - \delta_{\alpha, \alpha'_n} \right) \right]
,}
\end{equation}
\textcolor{black}{where $n_{\rm NNs}$ is the number of NNs of the anion, $\alpha$ represents the anion type, $\alpha'_n$ represents type of the $n$-th anion NN and $\delta_{\alpha, \alpha'_n}$ is the Kronecker delta.} The factor $c_\alpha$ is determined in a similar manner, but for cation.
\\
\indent To incorporate the strain effects, we use the Harrison rule~\cite{harrison1989electronic,Jancu1998,Harrison_rule1979} with $\eta=2$ \textcolor{black}{to keep the model physically as simple as possible associated with free-electron spectra~\cite{Jancu1998}} and multiply the off-site matrices by the factor representing an increase (decrease) of the hopping term reflecting a shortening (stretching) of the bond length $|{\bf t}|$ as 
%
\begin{equation}\label{99c}
\begin{split}
&H^\alpha_{{AC}}\cdot\left(\frac{|{\bf D}_\alpha|}{|{\bf t}|}\right)^\eta,\\
&H^\alpha_{{CA}}\cdot\left(\frac{|{\bf D}_\alpha|}{|{\bf t}|}\right)^\eta,
\end{split}
\end{equation}
where $\alpha\in\{\textup{Si-Si},\textup{C-C}\}$ enumerates different types of off-site parameters; $|{\bf D}_\beta|$ ($|{\bf t}|$) represents bulk unstrained (strained) bond distances between the anion and the cation.
%
%
%
%
%
%
%
%
\subsubsection{Approximations for on-site ETB matrices}
VCA modifies also on-site matrices as
\begin{equation}\label{VCA_on}
\begin{split}
&H_{{A}}=\frac{j_\gamma\cdot H_{{A}}^\gamma}{\sum_\gamma j_\gamma},\\
&H_{{C}}=\frac{j_\gamma\cdot H_{{C}}^\gamma}{\sum_\gamma j_\gamma},\\
&j_\gamma=\sum_n^{n_{\rm NNs}}~i_{\gamma}^n,
\end{split}
\end{equation}
where $n_{\rm NNs}$ is the number of NNs, $i_\gamma$ is taken from equation~\eqref{VCA_off} , $j_\gamma/\sum_\gamma j_\gamma$ represents the weights of using different types of on-site ETB parameters for anions or cations, $\gamma\in\{\textup{Si},\textup{C}\}$; Einstein sum notation is used in the first two equations in Eq.~\eqref{VCA_on}. 
%
%
\\
\indent Band offsets are incorporated into the model by adding the following terms to the on-site matrices along the main diagonal
\begin{equation}\label{offset}
\begin{split}
&H_{{A}}^\xi+\hat{I}\cdot\Delta E^{\rm offs.}_\xi ,\\
&H_{{C}}^\xi+\hat{I}\cdot\Delta E^{\rm offs.}_\xi ,
\end{split}
\end{equation}
where $\hat{I}$ is the identity matrix and $\xi\in\{\textup{Si},\textup{C}\}$. 
\subsubsection{Definition of the wavefunction in ETB model}
For a certain ${\bf K}$-vector from the first Brillouin zone (FBZ) of the supercell determined by the Born-von K\'arm\'an boundary condition, we diagonalize the Hamiltonian matrix (see Fig.~\ref{fig:Ham}) and obtain eigenenergies $E_{p,\bf K}$ and eigenvectors,~i.e. variation coefficients $b_{n,\alpha}^{ \upsilon,\sigma}(p,\mathbf{K})$ in the wavefunction of SC which is defined as
\begin{equation}\label{psi_SC}
\begin{split}
&|{\Psi^\textup{SC}_p(\mathbf{K})}\rangle=\sum^N_{n,\alpha,\upsilon,\sigma} e^{\textrm{i}\mathbf{K}\cdot (\mathbf{R}_{n}+\mathbf{d_\alpha})} \;b_{n,\alpha}^{ \upsilon,\sigma}(p,\mathbf{K})\;|\mathbf{R}_{n},\alpha,\upsilon,\sigma\rangle,\\
&\;\;\;\;\;\;\;\;\;\;\;\;\;\;\;\;\;\;\;\;\;\;\textrm{with}~\sum^N_{n,\alpha,\upsilon,\sigma}|b_{n,\alpha}^{ \upsilon,\sigma}(p,\mathbf{K})|^2=1,
\end{split}
\end{equation}
where $p$ labels the eigenstates of SC and $n$, $\alpha$, $\upsilon$, $\sigma$ represent the position of PC within SC, the position within PC as well as the type of atom, the atomic orbital and the spin, respectively. Hence, $N=N_{PC}N_{\alpha,\nu,\sigma}$ represents the product of the number of PCs in the SC (labeled by $N_{PC}$) with the number of atoms in PC multiplied by the number of orbitals
of electrons with opposite spin (labeled by $N_{\alpha,\nu,\sigma}$ which also represents the number of PC basis bands). The wavefunction with coefficients $c_{\alpha}^{ \upsilon,\sigma}(q,\mathbf{K})$ for just one PC at the position $\mathbf{R}_{n}$ is
\begin{equation}\label{psi_PC}
\begin{split}
&|{\psi^\textup{PC}_q(\bf K}),\mathbf{R}_{n}\rangle=e^{\textrm{i}\mathbf{\mathbf{K}\cdot R_{n}}}~\sum^{N_{\alpha,\upsilon,\sigma}}_{\alpha,\upsilon,\sigma} e^{\textrm{i}\mathbf{K}\cdot \mathbf{d_\alpha}} \;c_{\alpha}^{ \upsilon,\sigma}(q,\mathbf{K})\;|\mathbf{R}_{n},\alpha,\upsilon,\sigma\rangle,\\
&\;\;\;\;\;\;\;\;\;\;\;\;\;\;\;\;\;\;\;\;\;\;\textrm{with}~\sum^{N_{\alpha,\upsilon,\sigma}}_{\alpha,\upsilon,\sigma}|c_{\alpha}^{ \upsilon,\sigma}(q,\mathbf{K})|^2=1,
\end{split}
\end{equation}
where $q$ labels the eigenstates of PC. For one PC, $\mathbf{R}_{n}$ is set to zero.
\subsection{Band unfolding}
\subsubsection{Unfolding for bulk}
%
\begin{figure*}[htbp]
    \includegraphics[width=170mm]{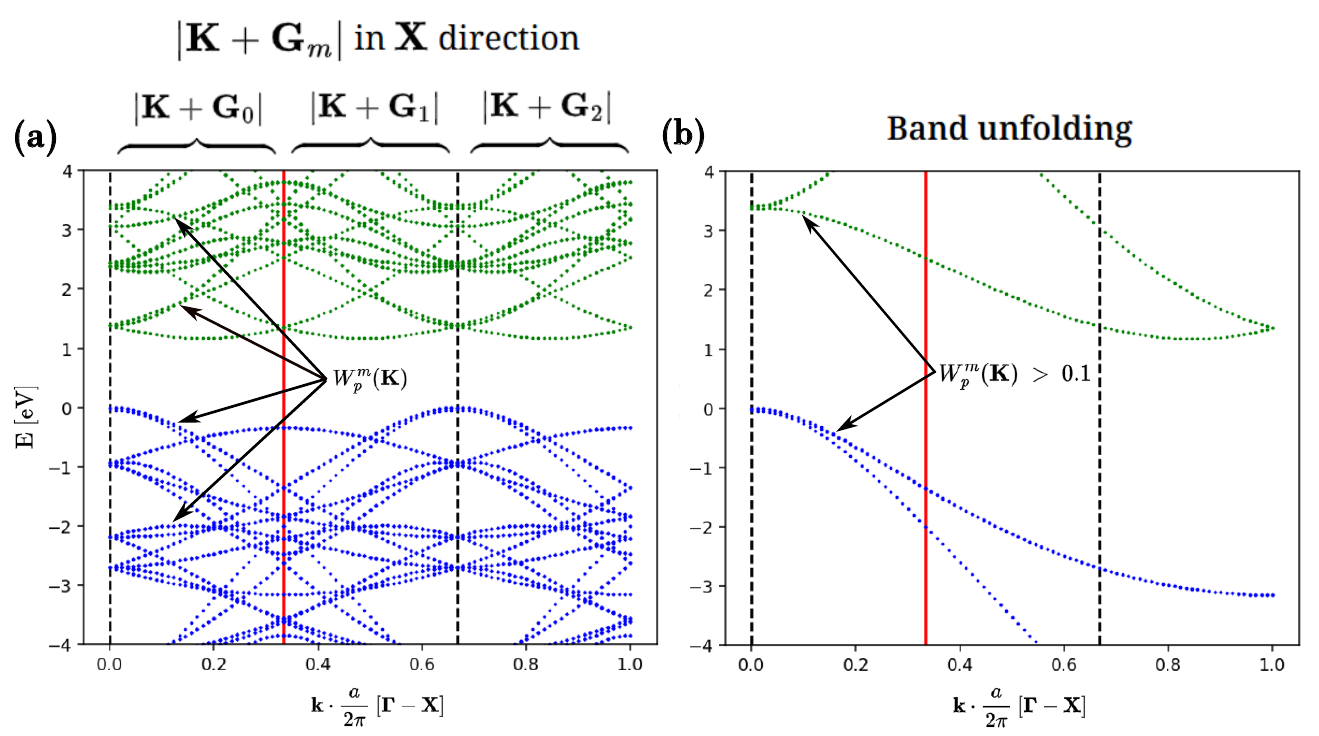}
	\caption{(a) Folded band structure of the silicon SC composed from $3\times 3\times3$ PCs with $|\mathbf{k}|=|\mathbf{K}+\mathbf{G}_{m}|$. There are only three reciprocal vectors $\mathbf{G}_{m}$ (out of total number of 27 reciprocal vectors) pointing in $\mathbf{X}$ direction ($\mathbf{G}_{0}=[0,0,0],~\mathbf{G}_{1}=[-0.66,0,0],~\mathbf{G}_{2}=[0.66,0,0]$). The weights $W_p^m(\mathbf{K})$ are calculated for each $m$-th reciprocal vector and for all eigenstates $p$ of the SC. The black broken vertical lines in (a) and (b) represent the absolute values of the three reciprocal vectors and red lines represent the edge of the FBZ of the SC. (b) Unfolded band structure, plotted for states with $W_p^m(\mathbf{K})>0.1$ only. }
	\label{fig:folded_Si}
\end{figure*}    
%
The band unfolding is a useful approach for bulk, alloys and nanostructures represented by a large supercell and,~thus, having a folded band structure of its Brillouin zone. 
\\
\indent The unfolding utilizes the linear approximation for the eigenfunctions of SC $|{\Psi^\textup{SC}_p(\mathbf{K})}\rangle$ in Eq.~\eqref{psi_SC}.  We introduce a new basis for SC $|{\psi^\textup{IFT~of~PC}_q(\mathbf{K}),\mathbf{G}_{m}}\rangle$ defined as an inverse Fourier transformation (IFT) of PC eigenstates $|\psi^\textup{PC}_q(\mathbf{K}),\mathbf{R}_{n}\rangle$ in Eq.~\eqref{psi_PC}, leading to
%
%
\begin{equation}\label{psi_IFTofPC}
\begin{split}
&|{\psi^\textrm{IFT~of~PC}_q(\mathbf{K}),\mathbf{G}_{m}}\rangle=\frac{1}{\sqrt{N_{PC}}}\sum^{N_{PC}}_{n }~e^{\textrm{i}\mathbf{\mathbf{G}_{m}\cdot R_{n}}}~|{\psi^\textup{PC}_q(\bf K}),\mathbf{R}_{n}\rangle\\
&=\frac{1}{\sqrt{N_{PC}}}\sum^{N_{PC}}_{n } e^{\textrm{i}\mathbf{(\mathbf{K}+\mathbf{G}_{m})\cdot R_{n}}} \sum^{N_{\alpha,\upsilon,\sigma}}_{\alpha,\upsilon,\sigma}\;e^{\textrm{i}\mathbf{\mathbf{K}\cdot d_{\alpha}}}c_{\alpha}^{\upsilon,\sigma}(q,\mathbf{K})\;|\mathbf{R}_{n},\alpha,\upsilon,\sigma\rangle,\\
&\textrm{with}~\sum^{N_{\alpha,\upsilon,\sigma}}_{\alpha,\upsilon,\sigma}|c_{\alpha}^{ \upsilon,\sigma}(q,\mathbf{K})|^2=1
\end{split}
\end{equation}
%
where the factor ${1}/{\sqrt{N_{PC}}}$ represents the normalization condition and the reciprocal vector $\mathbf{G}_{m}$ is from the FBZ of one PC (specified below). Notice that a new basis $|{\psi^\textup{IFT~of~PC}_q(\mathbf{K}),\mathbf{G}_{m}}\rangle$ and the old one $|\mathbf{R}_{n},\alpha,\upsilon,\sigma\rangle$ span the same space.~\cite{Boykin2016} The eigenfunctions of SC may be written as a linear combination in the new basis of inverse Fourier transformation of PC as~\cite{Boykin2016}          
%
%
\begin{equation}\label{unf_psi}
\begin{split}
&|{\Psi^\textup{SC}_p(\mathbf{K})}\rangle=\sum_{m}^{N_{PC}} \sum_{q}^{N_{\alpha,\upsilon,\sigma}}  \;a^m_{q}(p,\mathbf{K})\;|{\psi^\textrm{IFT~of~PC}_q(\mathbf{K}),\mathbf{G}_{m}}\rangle,\\
&\textrm{with}~\sum_{m}^{N_{PC}} \underbrace{\sum_{q}^{N_{\alpha,\upsilon,\sigma}}~|a^m_{q}(p,\mathbf{K})|^2}_{W_p^m(\mathbf{K})}=\sum_{m}^{N_{PC}}~W_p^m(\mathbf{K})=1,
\end{split}
\end{equation}
where $q$ labels the eigenstates of PC and $m$ goes through reciprocal wavevectors $\mathbf{G}_{m}$. The weight $W_p^m(\mathbf{K})$ represents the probability that SC eigenstate $|{\Psi^\textup{SC}_p(\mathbf{K})}\rangle$ has the same periodicity and behavior as bulk (described by one periodically repeating PC) with the wavevector $|\mathbf{k}|=|\mathbf{K}+\mathbf{G}_{m}|$ in FBZ of one PC. 
\\
\indent By inserting~\eqref{psi_SC} and~\eqref{psi_IFTofPC} in~\eqref{unf_psi} and writing the result in matrix form for $n$-th PC and $\alpha$-th atom in PC and orbitals with spin labeled by \textcolor{black}{$\omega=(\upsilon,\sigma)$} for simplification,
we get~\cite{Boykin2016}
%
%
\begin{equation}\label{unf_mat}
\begin{split}
&\underbrace{b_{n,\alpha}^{\omega}(p,\mathbf{K})}_{[\mathbf{B}^{\omega,\alpha}_{p}(\mathbf{K})]_n}=\sum_{m}^{N_{PC}}\underbrace{\frac{1}{\sqrt{N_{PC}}}e^{\textrm{i}\mathbf{\mathbf{G}_{m}\cdot R_{n}}}}_{[\mathbf{U}]_{n,m}}~\underbrace{\sum_{q}^{N_{\alpha,\upsilon,\sigma}}a^m_{q}(p,\mathbf{K})~c_{\alpha}^{\omega}(q,\mathbf{K})}_{[\mathbf{C}^{\omega,\alpha}_{p}(\mathbf{K})]_m},\\
&[\mathbf{B}^{\omega,\alpha}_{p}(\mathbf{K})]_n=[\mathbf{U}]_{n,m}\cdot[\mathbf{C}^{\omega,\alpha}_{p}(\mathbf{K})]_m.
\end{split}
\end{equation}
%
The matrix $[\mathbf{C}^{\omega,\alpha}_{p}(\mathbf{K})]_m$ represents the discrete Fourier transformation of the matrix $[\mathbf{B}^{\omega,\alpha}_{p}(\mathbf{K})]_n$ and recalling that the $N_{PC}\times N_{PC}$ matrix $[\mathbf{U}]_{n,m}$ is unitary, the solution can be readily obtained as~\cite{Boykin2016}    
\begin{equation}\label{unf_C_mat}
\begin{split}
&[\mathbf{C}^{\omega,\alpha}_{p}(\mathbf{K})]_m=[\mathbf{U}]^{\dagger}_{m,n}\cdot[\mathbf{B}^{\omega,\alpha}_{p}(\mathbf{K})]_n\\
&[\mathbf{C}^{\omega,\alpha}_{p}(\mathbf{K})]_m=\sum_{n}^{N_{PC}}~\frac{1}{\sqrt{N_{PC}}}e^{\textrm{-i}\mathbf{\mathbf{G}_{m}\cdot R_{n}}}~b_{n,\alpha}^{\omega}(p,\mathbf{K}).
\end{split}
\end{equation}
The weight $W_p^m(\mathbf{K})$ in Eq.~\eqref{unf_psi} can then be found as~\cite{Boykin2016} 
%
%
\begin{equation}\label{unf_weight}
\begin{split}
W_p^m(\mathbf{K})&=\sum_{q}^{N_{\alpha,\upsilon,\sigma}}|a^m_{q}(p,\mathbf{K})|^2=
\sum_{\omega,\alpha}^{N_{\alpha,\upsilon,\sigma}}|[\mathbf{C}^{\omega,\alpha}_{p}(\mathbf{K})]_m|^2\\
&=\sum_{\omega,\alpha}^{N_{\alpha,\upsilon,\sigma}}|[\mathbf{U}]^{\dagger}_{m,n}\cdot[\mathbf{B}^{\omega,\alpha}_{p}(\mathbf{K})]_n|^2\\
&\;\;\textrm{with}~\sum^{N}_{p}W_p^m(\mathbf{K})=N_{\alpha,\upsilon,\sigma},
\end{split}
\end{equation}
where we added $\sum_{\alpha,\upsilon,\sigma}|c_{\alpha}^{ \upsilon,\sigma}(p,\mathbf{K})|^2=1$ from~\eqref{psi_IFTofPC} in the second step. 
\\
\indent To calculate weights $W_p^m(\mathbf{K})$  from ETB coefficients $[\mathbf{B}^{\omega,\alpha}_{p}(\mathbf{K})]_n$, we need to find the vectors $\mathbf{G}_{m}$ from which $[\mathbf{U}]_{n,m}$ can be built. Those are obtained from the number of PCs $N_1,N_2, N_3$ that create the SC in the direction of primitive lattice vectors ${\bf a}_1,{\bf a}_2,{\bf a}_3$, respectively. Vectors $\mathbf{G}_{m}$ are thus
\begin{equation}\label{unf_G_vect}
\begin{split}
&\mathbf{G}_{m}=\sum_i^3~\frac{m_i}{{N_{i}}}\mathbf{b}_{i},
\end{split}
\end{equation}
where $\mathbf{b}_{i}$ are the PC reciprocal lattice vectors and $m_i=\{-(N_i-2)/2,~...,~0,~...,N_i/2\}$ for even $N_i$, and $m_i=\{-(N_i-1)/2,~...,~0,~...,(N_i-1)/2\}$ for odd $N_i$.~\cite{Boykin2016} 
\\
\indent The unfolding process is illustrated in Fig.~\ref{fig:folded_Si}. From Eq.~\eqref{unf_weight}, we calculate the weights $W_p^m(\mathbf{K})$ for all SC eigenstates and all reciprocal vectors $\mathbf{G}_{m}$. Finally, we consider only weights for which $\mathbf{G}_{m}$ vectors point to the required direction.
\subsubsection{Unfolding for the G-center}
In the case of the G-center, SC is no longer fully composed of PCs, and so the approach above must be modified. The coefficients $b_{n,\alpha}^{\omega}(p,\mathbf{K})$ of SC need to be separated into the periodic part (which belongs to the bulk, labeled by $b_{n_\textrm{bk},\alpha_\textrm{bk}}^{'\omega}(p,\mathbf{K})$) and the aperiodic part (which belongs to the G-center, divided into parts belonging to the substitutional $b_{n_\textrm{sub},\alpha_\textrm{sub}}^{''\omega}(p,\mathbf{K})$ and interstitial $b_{n_\textrm{int},\alpha_\textrm{int}}^{''\omega}(p,\mathbf{K})$ atoms). Eq.~\eqref{unf_weight} is then modified as
%
%
\begin{equation}\label{unf_W_mat_mod}
\begin{split}
&[\tilde{\mathbf{B}}^{\omega,\alpha}_{p}(\mathbf{K})]_n:=[\underbrace{b_{n_\textrm{bk},\alpha_\textrm{bk}}^{'\omega}(p,\mathbf{K})}_{\textrm{periodic~part}},~\underbrace{0\cdot b_{n_\textrm{sub}, \alpha_\textrm{sub}}^{''\omega}(p,\mathbf{K})}_{\textrm{aperiodic~part}}]\\
&\tilde{W}_p^m(\mathbf{K})=
\sum_{\omega,\alpha}^{N_{\alpha,\upsilon,\sigma}}|[\mathbf{U}]^{\dagger}_{m,n}\cdot[\tilde{\mathbf{B}}^{\omega,\alpha}_{p}(\mathbf{K})]_n|^2.
\end{split}
\end{equation}
We further separate the contribution of the aperiodic part corresponding to the interstitial atom and that corresponding to the substitutional atoms. Although the former is excluded from Eq.~\eqref{unf_W_mat_mod} altogether, the coefficients of the latter are multiplied by zero. In that way, the net effect of the aperiodic parts is excluded from the unfolding procedure, but the corresponding matrix rank is kept because of the zeroing of $b_{n_\textrm{sub}, \alpha_\textrm{sub}}^{''\omega}(p,\mathbf{K})$.
%
%
The $N_{PC}$ then contains primitive cells originating from bulk and that due to the substitutional atoms. 
\\
\indent The assumptions in Eqs.~\eqref{unf_weight} and~\eqref{unf_psi} no longer apply and need to be changed to
\begin{equation}\label{unf_mod_assum}
\begin{split}
&\sum^{N}_{p}\tilde{W}_p^m(\mathbf{K})=N_{\alpha,\upsilon,\sigma}-N_{\alpha,\upsilon,\sigma}/N_{PC},\\
&\sum_{m}^{N_{PC}}~\tilde{W}_p^m(\mathbf{K})=1-\underbrace{\sum_{\omega,\alpha_\textrm{int}}^{N_{\alpha_\textrm{int},\upsilon,\sigma}}|b_{n_\textrm{int},\alpha_\textrm{int}}^{''\omega}(p,\mathbf{K})|^2}_{\textrm{int. atoms}}\\
&~~~~~~~~~~~~~~~~~~~~~~~-\underbrace{\sum_{\omega,\alpha_\textrm{sub}}^{N_{\alpha_\textrm{sub},\upsilon,\sigma}}|b_{n_\textrm{sub},\alpha_\textrm{sub}}^{''\omega}(p,\mathbf{K})|^2}_{\textrm{sub. atoms}}.
\end{split}
\end{equation}
The total weight of SC of the G-center $W_p^m(\mathbf{K})$ is then modified as
\begin{equation}\label{unf_weight_last}
\begin{split}
&W_p^m(\mathbf{K})=\tilde{W}_p^m(\mathbf{K})+\overbrace{W^\textrm{int}_p(\mathbf{K})+W^\textrm{sub}_p(\mathbf{K})}^{W^\textrm{G-center}_p(\mathbf{K})}\\
&=
\underbrace{\sum_{\omega,\alpha}^{N_{\alpha,\upsilon,\sigma}}|[\mathbf{U}]^{\dagger}_{m,n}\cdot[\tilde{\mathbf{B}}^{\omega,\alpha}_{p}(\mathbf{K})]_n|^2}_{\textrm{bulk}}\\
&+\underbrace{\sum_{\omega,\alpha_\textrm{int}}^{N_{\alpha_\textrm{int},\upsilon,\sigma}}|b_{n_\textrm{int},\alpha_\textrm{int}}^{''\omega}(p,\mathbf{K})|^2}_{\textrm{int. atoms}}
+\underbrace{\sum_{\omega,\alpha_\textrm{sub}}^{N_{\alpha_\textrm{sub},\upsilon,\sigma}}|b_{n_\textrm{sub},\alpha_\textrm{sub}}^{''\omega}(p,\mathbf{K})|^2}_{\textrm{sub. atoms}},
\end{split}
\end{equation}
where $|b_{n_\textrm{int},\alpha_\textrm{int}}^{''\omega}(p,\mathbf{K})|^2$ and $|b_{n_\textrm{sub},\alpha_\textrm{sub}}^{''\omega}(p,\mathbf{K})|^2$ do not depend on index $m$ and as a result the weights representing the G-center $W^\textrm{int}_p(\mathbf{K})$ and $W^\textrm{sub}_p(\mathbf{K})$ do not depend on $m$ as well. 
%
%
\begin{figure*}[htbp!]
    \includegraphics[width=170mm]{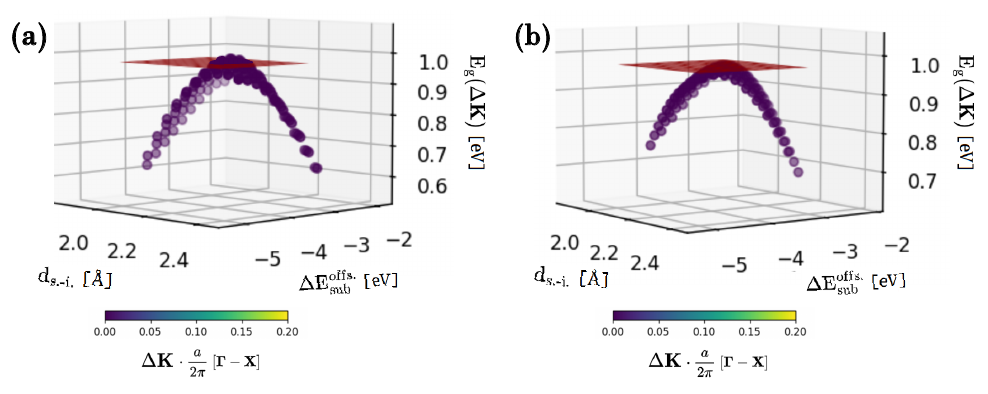}
	\caption{Energy band gap of the G-center (for i=0 and $N_{SC}=5$) with a dependence on the band offset of the substitutional atom $\Delta {E}^{\rm offs.}_\textrm{sub}$ and the bond length ${d}_{\textrm{s.-i.}}$ where the color bar marks a difference of reciprocal wavevector during the transition (dark balls represent direct transition). Red plane indicates the energy of the G-center of $970$ meV. Panel (a) gives the case without and (b) with adding the band offset of interstitial atom $\Delta {E}^{\rm offs.}_\textrm{int}=0.63$~eV which lowers the maximum of band gap of G-center transition energy.}
	\label{fig:map_555}
\end{figure*}
%
\subsection{Configuration Interaction}
The configuration Interaction (CI) adds the multi-particle Coulomb interaction correction to the single-particle model. We first introduce multi-particle basis states as the Slater determinants (SDs).~\cite{Bryant1987,Troparevsky2008,Schliwa:09,Klenovsky2017,Huang2021a} For neutral exciton X, the SDs are written as
\begin{equation}\label{CI_base}
\begin{split}
|SD^\textrm{X}\rangle&=\sum_i^{n_h}\sum_j^{n_e}\underbrace{\frac{1}{\sqrt{2}}\begin{vmatrix}\Psi_i(\mathbf{r}_e) & \Psi_i(\mathbf{r}_h) \\ \Psi_j(\mathbf{r}_e) & \Psi_j(\mathbf{r}_h)  \end{vmatrix}}_{|SD^\textrm{X}_{ij}\rangle}=\sum_m^{n_en_h}|SD^\textrm{X}_m\rangle
\end{split}
\end{equation}
where $n_e$ ($n_h$) represents the number of single-particle states of electrons (holes) at positions ${\bf r}_e$ (${\bf r}_h$) that make up the $m$-th SD $|SD^\textrm{X}_m\rangle$. 
\\
\indent The multi-particle trial wavefunction is then considered as a linear combination of SDs
\begin{equation}\label{CI_state}
\begin{split}
&\Phi^\textrm{X}_s(\mathbf{r}_a,\mathbf{r}_b)=\sum^{n_en_h}_m ~\eta^m_{s}~|SD^\textrm{X}_m\rangle,\\
&\textrm{with}~\sum^{n_en_h}_m|\eta^m_{s}|^2=1,
\end{split}
\end{equation}
where $s$ enumerates the multi-particle eigenstate $\eta^m_{s}$ of the corresponding Schrödinger equation. The CI Hamiltonian is obtained as
%
%
\begin{equation}\label{CI_ham}
\begin{split}
&\hat{H}_{mn}^\textrm{X}=\delta_{mn}(E_{e(j)}-E_{h{(i)}})_m+\langle SD^\textrm{X}_m|\hat{V}_{eh}|SD^\textrm{X}_n\rangle\\
&\textrm{with}~\hat{V}_{eh}=-\frac{e^2}{4\pi\epsilon_0\epsilon_{r}(\mathbf{r}_e, \mathbf{r}_h)}\frac{1}{|\mathbf{r}_e-\mathbf{r}_h|},
\end{split}
\end{equation}
where the minus sign at the beginning corresponds to the exciton. 
\\
\indent In the case of $n_e=2$ and $n_h=2$, where $m,n=\{1,3;2,3;1,4;2,4\}$ with $i,i'=\{1,2\}$ and $j,j'=\{3,4\}$, the matrix elements $\langle SD^\textrm{X}_m|\hat{V}_{eh}|SD^\textrm{X}_n\rangle$ are obtained \{see Eq.~\eqref{CI_base}\} as
%
\begin{equation}\label{CI_inter}
\begin{split}
&\langle SD^\textrm{X}_m|\hat{V}_{eh}|SD^\textrm{X}_n\rangle=\langle SD^\textrm{X}_{i'j'}|\hat{V}_{eh}|SD^\textrm{X}_{ij}\rangle=\\
&\frac{1}{2}\sum_{e}^{N_{n,\alpha}}\sum_{h,~h\neq e}^{N_{n,\alpha}}\frac{1}{2}\{\hat{V}_{eh}\Psi^*_{i'}(\mathbf{r}_e)\Psi^*_{j'}(\mathbf{r}_h)\Psi_i(\mathbf{r}_e)\Psi_j(\mathbf{r}_h)\\
&+\hat{V}_{eh}\Psi^*_{i'}(\mathbf{r}_h)\Psi^*_{j'}(\mathbf{r}_e)\Psi_i(\mathbf{r}_h)\Psi_j(\mathbf{r}_e)\\
&-\hat{V}_{eh}\Psi^*_{i'}(\mathbf{r}_e)\Psi^*_{j'}(\mathbf{r}_h)\Psi_i(\mathbf{r}_h)\Psi_j(\mathbf{r}_e)\\
&-\hat{V}_{eh}\Psi^*_{i'}(\mathbf{r}_h)\Psi^*_{j'}(\mathbf{r}_e)\Psi_i(\mathbf{r}_e)\Psi_j(\mathbf{r}_h)\},
\end{split}
\end{equation}
%
%
where \textcolor{black}{$N_{n,\alpha}$ is the number of all atoms in the structure}, the factor $\frac{1}{2}$ before sums prevents double counting and factor $\frac{1}{2}$ after the sums is derived from bra-ket of SDs in Eq.~\eqref{CI_base}. 
The CI Hamiltonian in Eq.~\eqref{CI_ham} then forms the $n_en_h\times n_en_h$,~i.e. \textcolor{black}{$4\times 4$} matrix which has to be diagonalized and we obtain the CI corrected band gap energies for the exciton. 
\\
\indent Furthermore, note that the summations over atom positions in Eq.~\eqref{CI_inter} for ETB wavefunctions replace integration in continuum methods like,~e.g., ${\bf k}\cdot{\bf p}$ approximation.~\cite{Klenovsky2017}
\section{Results}
\label{sec:expTeorCompar}
%
\begin{figure*}[htbp]
    \includegraphics[width=170mm]{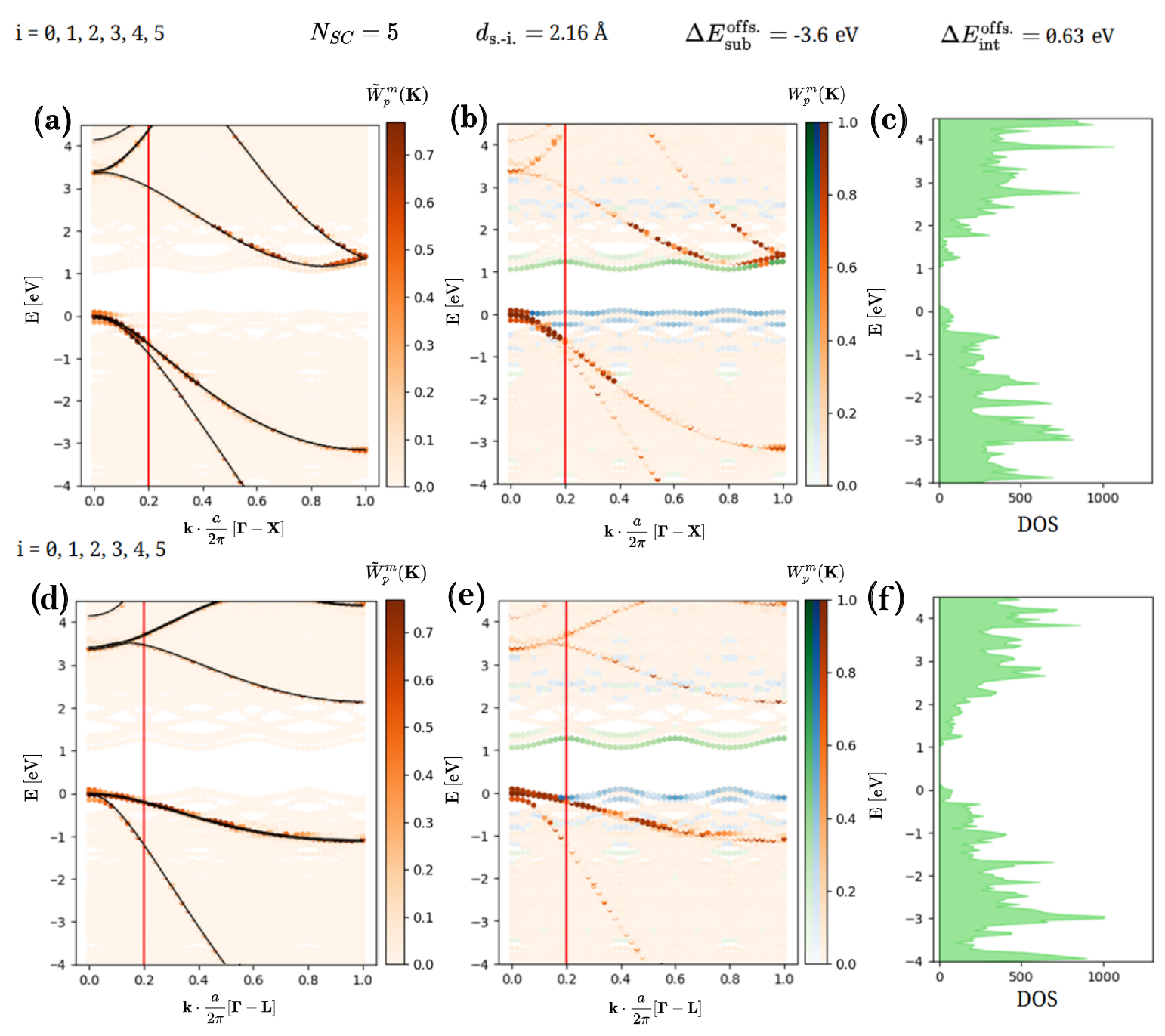}
	\caption{Unfolded band structure of the G-center (bandstructures are the same for all i=0-5) with $|\mathbf{k}|=|\mathbf{K}+\mathbf{G}_{m}|$ and the calculated density of states (DOS). The transition energy is $E_\textrm{g}(\Delta\mathbf{K})=970.01$ meV. The values of external tuning parameters for which the band structure was calculated are given in the inset at the top of the graph. (a)-(c) electronic band structure from $\mathbf{\Gamma}$ to $\mathbf{X}$ ${\bf k}$-point and (d)-(f) that from $\mathbf{\Gamma}$ to $\mathbf{L}$ ${\bf k}$-point. The color bars represent the weight ${W}_p^m(\mathbf{K})$. The red color bar in (a) and (d) represents the weight of pure bulk $\tilde{W}_p^m(\mathbf{K})$ and black thin curve follows band structure of silicon bulk calculated by ETB for one PC. In (b) and (e) the color bar for ${W}_p^m(\mathbf{K})$ is given in three colors according to main contributions in sum of Eq.~\eqref{unf_weight_last} where red stands for bulk, blue for interstitial silicon atom and green for substitutional carbon atoms. The red vertical lines represent the edge of the FBZ of the SC. In (c) \{(f)\} DOS computed from unfolded band structure in (b) \{(e)\} is shown.}
	\label{fig:Gc_555}
\end{figure*}
%
\begin{figure*}[htbp]
    \includegraphics[width=170mm]{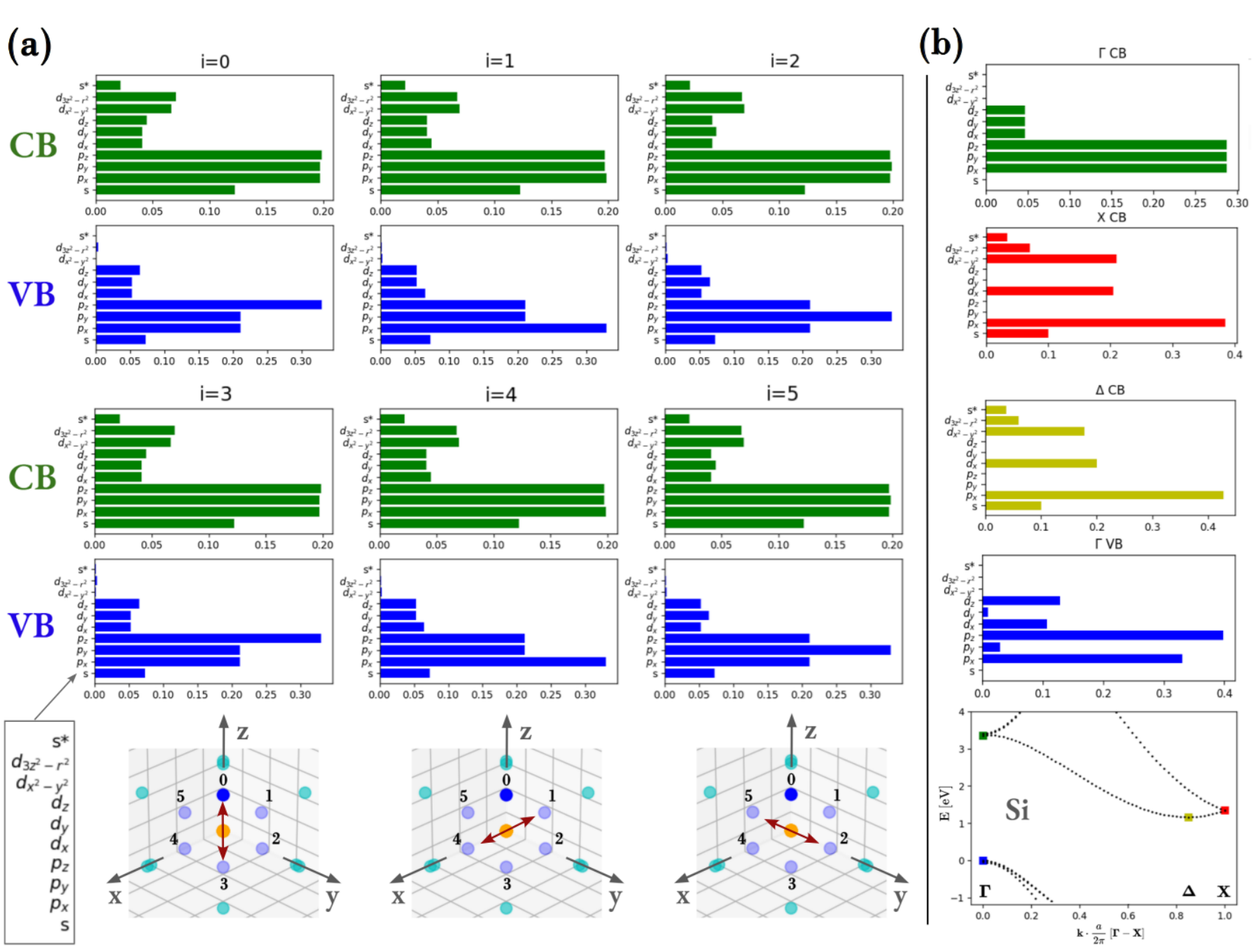}
	\caption{(a) Orbital decomposition of sp$^3$d$^5$s* bases for electron-hole transition of the G-center at $\mathbf{\Gamma}$-point for the interstitial configurations i=0-5 with band gap energy $E_\textrm{g}(\Delta\mathbf{K})=970.01$ meV. The three distinct silicon interstitial configurations are identified in each column of (a) with the atomic positions \{see also Fig.~\ref{fig:Struct}~(a)\} given at the bottom of the columns. (b) Orbital decomposition of sp$^3$d$^5$s* bases for silicon bulk for $\mathbf{\Gamma}$-, $\mathbf{X}$-, and $\mathbf{\Delta}$-points of band structure depicted at the bottom of panel (b).}
	\label{fig:Orb_dec}
\end{figure*}
%
\subsubsection{\textcolor{black}{Calculation of the electronic band structures of the G-center}}
We calculated the electronic band structures of the G-center for a dependence on three parameters: (i) the bond length $d_{\textrm{s.-i.}}$ between the substitutional carbon and the interstitial silicon atoms in the G-center around 2~\AA~\cite{Gc_bond_length2014}, (ii) the band offset $\Delta E^{\rm offs.}_\textrm{sub}$ added to substitutional carbon on-site ETB parameters, see Eq.~\eqref{offset}, and (iii) the size of the SC $N_{SC}=N_1=N_2=N_3$. 
The reason for introducing these parameters is that ETB bulk parameters in the presence of the G-center must be modified because the G-center no longer behaves as a bulk material. The bond length and the SC size are natural parameters that must be specified before the ETB calculation and are introduced with the definition of SC itself. However, the band offset represents the requirement of the change of the on-site parameters to correctly describe the electron-hole transition within the G-center. 
\\
\indent The energy band gap between the conduction band (CB) and the valence band (VB) was established between two bands labeled $2\mathcal{N}_{CB}$ and $2\mathcal{N}_{VB}$ (the factor two is added due to spin degeneracy) as 
\begin{equation}\label{Eg_set}
\begin{split}
&E_\textrm{g}(\Delta\mathbf{K})=E_{2\mathcal{N}_{CB}}(\mathbf{K}_{2\mathcal{N}_{CB}})-E_{2\mathcal{N}_{VB}}(\mathbf{K}_{2\mathcal{N}_{VB}})\\
&2\mathcal{N}_{VB}=(N_\textrm{all~at.}\times 4-2)\\
&2\mathcal{N}_{CB}=(N_\textrm{all~at.}\times 4-2)+2,
\end{split}
\end{equation}
where $\Delta\mathbf{K}=|\mathbf{K}_{2\mathcal{N}_{CB}}-\mathbf{K}_{2\mathcal{N}_{VB}}|$, $N_\textrm{all~at.}$ is the number of all atoms in SC (bulk + sub. + int.), 4 represents the number of valence electrons in silicon and the subtraction by 2 is due to two dangling interstitial bonds in the model, which are not used in the calculation. We are aware that an interstitial silicon atom has four valence electrons and, hence, as a simplification of the problem two rest valence electrons are not considered in the calculation. 
\textcolor{black}{For the definition of the minimum of the CB above the maximum of the VB, a value of 2 is added as a summand in the last equation in~ \eqref{Eg_set}. In the present calculation, we consider only spin-conserving transitions (i.e., without spin flip processes). }
\subsubsection{\textcolor{black}{Band gap tuning of the G-center via three key parameters}}
\indent According to the aforementioned three parameters (i)-(iii), the best agreement with the band gap energy of the G-center around $970$ meV with the ${\bf k}$-direct transition~\cite{Gc_Durand2024_main,Gc_B_cof_Shainline2007,Gc_Beaufils2018_navic,Gc_OSC_2type_II,Density_of_Gc2011,Gc_Durand2024_OSC} is found for the case $N_{SC}=5$. For comparison, we have calculated also 3D maps for different sizes of the SC, see Appendix~I. 
In Fig.~\ref{fig:map_555}, we show the dependencies of the band gap energies ${E}_\textrm{g}(\Delta\mathbf{K})$ on the band offset $\Delta {E}^{\rm offs.}_\textrm{sub}$ and the bond length ${d}_{\textrm{s.-i.}}$ for a case of the position of the interstitial atom i=0. To simplify the problem, we choose the maximum band gap magnitude among many possible solutions.
%
%
The maximum band gap is found to have a value of $E_\textrm{g}(\Delta\mathbf{K})=981.99$~meV and occurs for $\Delta {E}^{\rm offs.}_\textrm{sub}=-3.6$ eV and ${d}_{\textrm{s.-i.}}= 2.16$~\AA, see Fig.~\ref{fig:map_555}~(a). 
\\
\indent Further, we include the band offset $\Delta {E}^{\rm offs.}_\textrm{int}=0.63$~eV for interstitial silicon as a next external tuning parameter, which lowers the maximum band gap to $E_\textrm{g}(\Delta\mathbf{K})=970.01$~meV, see Fig.~\ref{fig:map_555} (b). By doing so we were able to achieve an agreement of the calculated result with the measured band-gap energy of the G-center.~\cite{Gc_Durand2024_main,Gc_B_cof_Shainline2007,Gc_Beaufils2018_navic,Gc_OSC_2type_II,Density_of_Gc2011,Gc_Durand2024_OSC} From the size of SC ($N_{SC}=5$) for which the agreement with experiment was achieved, we \textcolor{black}{can} calculate the expected density of the G-center in bulk as $20\times 10^{19}$~cm$^{-3}$ which is five times denser then that determined experimentally in the case of high-density G-centers\cite{Density_of_Gc2011,Gc_OSC_2type_II}, where a density of $4\times 10^{19}$~cm$^{-3}$ was obtained. We note that while the agreement between theory and experiment is not precise, the G-center concentration obtained from theory is still remarkably close to that of the experiment \textcolor{black}{considering that the SC size is one of the tuning parameters, our estimate should be viewed as approximate, and the actual value may still differ}. 
\subsubsection{\textcolor{black}{Revealing the direct transition mechanism in the G-center via the band structure unfolding approach}}
\indent A detailed band structure for the case of $E_\textrm{g}(\Delta\mathbf{K})=970.01$~meV,~i.e. the maximal band gap in \{Fig.~\ref{fig:map_555} (b)\} is shown in Fig.~\ref{fig:Gc_555}.
Note that band structures for all interstitial positions i=0-5 look very similar, hence, only that for i=0 is given in Fig.~\ref{fig:Gc_555}. The color bar representing the total weight ${W}_p^m(\mathbf{K})$ in (b) and (e) has three colors according to main contribution in sum of Eq.~\eqref{unf_weight_last} where red stands for bulk, blue for interstitial silicon atom and green for substitutional carbon atoms. The calculated band gap transition is $\bf k$-direct at $\mathbf{\Gamma}$-point for all i=0-5 where the main contribution for direct transition is due to carbon atoms (green weights) as well as the interstitial atoms (blue weights), in Fig.~\ref{fig:Gc_555} (b) and (e). Notice that the band unfolding weight $\tilde{W}_p^m(\mathbf{K})$ of bulk in Fig.~\ref{fig:Gc_555} (a) represents silicon bulk behavior (calculated by ETB for one PC) depicted by black curve. Hence, we can conclude that the G-center leads to the direct electron-hole transition in bulk silicon which would otherwise have indirect band gap. Remarkably, the ETB model is able to describe the behavior of the G-center in silicon bulk and by unfolding approach it can distinguish the contributions of different elements in SC which have impact on the final band structure. The results also indicate how the band structure of bulk silicon (described by one PC) changes in the case of the presence of the G-center. 
%
\subsubsection{\textcolor{black}{Orbital decomposition of the electron-hole transition of the G-center}}
\indent In Fig.~\ref{fig:Orb_dec}~(a) we see the orbital decomposition of the electron-hole transition of the G-center at $\mathbf{\Gamma}$-point calculated in Fig.~\ref{fig:Gc_555}~(b)~and~(e). Notice that the orbital decompositions are the same for opposite positions of the interstitial atoms in the G-center (0 and 3, 1 and 4, 2 and 5) following symmetry axes of the G-center, see at the bottom of Fig.~\ref{fig:Orb_dec}~(a). \textcolor{black}{In Fig.~\ref{fig:Orb_dec}~(b) we show orbital decomposition of the silicon bulk for three CB (at different points of Brillouin zone: $\mathbf{\Gamma}$, $\mathbf{\Delta}$ and $\mathbf{X}$) and one VB (at $\mathbf{\Gamma}$-point).} If we compare the orbital decomposition of the G-center to the silicon bulk, we notice that the CB of the G-center (green)\textcolor{black}{, in Fig.~\ref{fig:Orb_dec}~(a)}, and the CB of bulk at $\mathbf{\Delta}$-point (yellow) and $\mathbf{X}$-point (red)\textcolor{black}{, in Fig.~\ref{fig:Orb_dec}~(b)}, are composed of the orbitals ($d_{x^2-y^2}$, $d_{3z^2-r^2}$, $s^*$) indicating that the transition in the G-center behaves similarly to the transition in bulk but with the direct band gap. The d-orbitals are incorporated into the ETB model in order to correctly describe the indirect band gap in silicon bulk and from orbital decomposition of electron-hole transition in the G-center, we may notice that they are also crucial for description of the direct transition in the G-center.
\begin{figure}[htbp]
    \includegraphics[width=85mm]{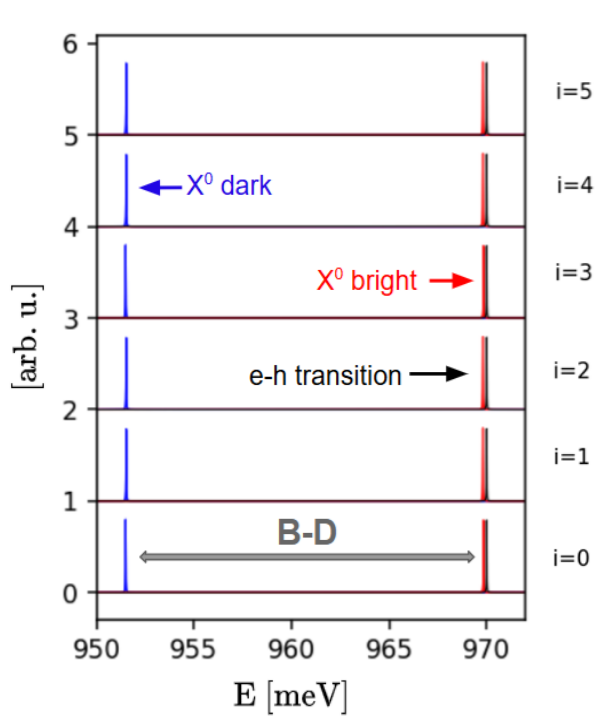}
	\caption{Exciton (X$^0$) energies for the G-centers with interstitial positions i=0-5 and electron-hole transition with the energy of $970.01$ meV. The X$^0$ states were computed with the CI basis of two ground state electron and two ground state hole wavefunctions. Notice that for all interstitial positions, the computed bright X$^0$ is very close to the experimentally observed transition in Refs.~\cite{Gc_Durand2024_main,Gc_B_cof_Shainline2007,Gc_Beaufils2018_navic,Gc_OSC_2type_II,Density_of_Gc2011,Gc_Durand2024_OSC} Moreover, notice the small binding energy of bright X$^0$ with respect to single-particle transition being $\approx160-200\,\mu$eV and much larger energy splitting of $\approx18$~meV between bright and dark X$^0$ states.}
	\label{fig:CI_555}
\end{figure}
\subsubsection{\textcolor{black}{Insights into the excitonic properties of the G-center via configuration interaction calculations}}
\indent The exciton correction of the electron-hole transition at $\mathbf{\Gamma}$ point computed by CI with the single-particle basis of two ground state electron and two ground state hole eigenstates for the band structure above was calculated for silicon interstitial position i=0-5 and the results are summarized in Table~\ref{tab:TB_CI_555} and shown in Fig.~\ref{fig:CI_555}. \textcolor{black}{The choice to consider two ground state electron and two ground state hole eigenstates was motivated by considering only states associated with the G-center that exhibit the highest weight (corresponding to the blue and green bands in Fig.~\ref{fig:Gc_555}).} Firstly, we notice that the CI predicts very small fine-structure splitting (FSS) of ground state exciton (X$^0$) both for bright and dark X$^0$ doublet. That suggests the silicon G-center to be potentially a very good emitter of entangled photons for quantum communication and computation applications.~\cite{Kimble2008,Fox2024,Liu2024,Rahmouni2024}
Interestingly, as noticed from Table~\ref{tab:TB_CI_555} the CI calculations show a slight difference (on the order of $\sim0.03\,\mu$eV) in FSS as well as X$^0$ energy between configurations i=\{0, 3\} and i=\{1, 2, 4, 5\}. While such a difference is very timid being on the order of numeric errors, it is clearly reproducible in the calculations. We speculate that this difference might originate in different orbital decomposition of the single-particle electron-hole transitions where electrons in configurations i=\{0, 3\} prefer to occupy \textcolor{black}{$d_{3z^2-r^2}$} orbitals (due to orientation of the interstitial atom in $z$ direction) but for i=\{1, 2, 4, 5\} they prefer to occupy $d_{x^2-y^2}$ orbitals (due to orientation of the interstitial atom in $x$ and $y$ directions). \textcolor{black}{Note that the orbitals ($d_{x^2-y^2}$, $d_{3z^2-r^2}$) responsible for the configuration asymmetry appear in the bulk band decomposition \{Fig.~\ref{fig:Orb_dec}~(b)\} at the $\mathbf{\Delta}$ and $\mathbf{X}$ points.} Moreover, notable on CI excitonic calculations of the G-center in silicon is a large splitting of bright and dark X$^0$ doublet (B-D) of approx. 18~meV,~i.e. five orders of magnitude larger than FSS. \textcolor{black}{We note that the dark exciton energy around 951.5~meV (1303~nm) may correspond to the PL signal measured at low temperatures (30~K) observed around 1300~nm in experiments on G* and, to a lesser extent, G-centers in Ref.~\cite{Gc_Durand2024_OSC}. However, our calculations do not predict the appearance of the E line.~\cite{Gc_Durand2024_OSC,Gc_Durand2024_main} Therefore, these theoretical studies of the G-center open up possibilities to explore structural modifications or incorporate strain to investigate other experimentally observed properties.} We also find a very shallow ($\approx 160-200\,\mu$eV) binding energy of bright X$^0$ when compared to the single-particle electron-hole transition. We finally note that the aforementioned very small FSS and large B-D splitting as well as the possibility to operate at elevated temperatures put the G-center in silicon in striking contrast to quantum dot (QD) sources currently studied for usage in quantum communication and computation.~\cite{Fox2024,Aharonovich2016,Senellart2017,zhou2023epitaxial,Rastelli2004}
%
\begin{table*}[htbp]
  \tabcolsep=10pt
  \centering\small
  \caption{Calculated dark fine structure splitting (FSS), bright FSS and bright-dark (B-D) energy separation of the ground state exciton X$^0$ for the G-centers with interstitial positions i=0-5. Moreover, for those \textcolor{black}{interstitial} position we give also the energy of the electron-hole single particle transition as well as that for bright and dark X$^0$. \textcolor{black}{We add orbital decomposition of $d_{3z^2-r^2}$ and $d_{x^2-y^2}$ in the CB and the VB to highlight the corresponding differences in Fig.~\ref{fig:Orb_dec}~(a).}}
  \vspace{0.5cm}
  \label{tab:TB_CI_555}
  \makebox[\textwidth]
  {\begin{tabular}{| c |c|c|c|c|c|c|}
      \hline \hline
        & i=0 & i=1 & i=2 & i=3 & i=4 & i=5 \\
        \hline  \hline
     FSS dark [$\mu$eV]&   0.655
&  0.141
&  0.142
&  0.655
&  0.142
&  0.141
\\
        \hline\hline
     FSS bright [$\mu$eV]& 0.336
& 0.234
& 0.234
& 0.336
& 0.234
&0.234
\\\hline
             \hline\hline
     B-D [meV]& 18.346& 18.288& 18.288& 18.346& 18.288&18.288\\\hline
             \hline\hline
     ETB e-h transition [meV]& 970.007& 970.007& 970.007& 970.007& 970.007&970.007\\\hline
             \hline\hline
     Bright exciton [meV]& 969.845& 969.815& 969.815& 969.845& 969.815&969.815\\\hline
 Dark exciton [meV]& 951.499& 951.527& 951.527& 951.499& 951.527&951.527\\\hline
             \hline\hline
 CB $d_{3z^2-r^2}$& 0.067& 0.065& 0.065& 0.067& 0.065&0.065\\\hline
 CB $d_{x^2-y^2}$& 0.064& 0.066& 0.066& 0.064& 0.066&0.066\\\hline
 VB $d_{3z^2-r^2}$& 0.003& 0.002& 0.002
& 0.003
& 0.002
&0.002
\\\hline
 VB $d_{x^2-y^2}$& 0.001& 0.003& 0.003& 0.001& 0.003&0.003\\\hline
            \hline\hline
    \end{tabular}
}
\end{table*}
%
\section{Discussion and conclusions}
\label{sec:conclusions}
In this work, we have theoretically studied the band structure of the silicon with the G-center using the empirical tight-binding model employing the unfolding approach and configuration interaction correction for excitons. 
We introduced into tight-binding three external tuning parameters to correctly describe the energy band gap of the G-center,~\cite{Gc_Durand2024_main,Gc_B_cof_Shainline2007,Gc_Beaufils2018_navic,Gc_OSC_2type_II,Density_of_Gc2011,Gc_Durand2024_OSC} i.e. the bond length of the G-center (between interstitial and substitutional atoms) and two band offsets (one for interstitial and one for substitutional atoms). 
%
The reason for introducing these two band offsets is that ETB bulk parameters in the presence of the G-center must be modified because the behavior of the G-center markedly departs from that of infinitely periodic bulk crystal. 
%
Using our approach, we have shown that the ${\bf k}$-direct electron-hole transition at ${\Gamma}$ point caused by adding the G-center to the silicon bulk, which under ambient conditions has indirect band gap, was possible. Moreover, the computed direct transition energy matches the measured band-gap energy of the G-center.~\cite{Gc_Durand2024_main,Gc_B_cof_Shainline2007,Gc_Beaufils2018_navic,Gc_OSC_2type_II,Density_of_Gc2011,Gc_Durand2024_OSC}.
Remarkably, the tight-binding model is able to describe not only the behavior of the G-center in the silicon bulk but using the unfolding approach it can also pinpoint the contributions of different elements of the supercell which create the final band structure. The results also indicate how the band structure of bulk silicon (described by one primitive cell) changes in the presence of the G-center. 
%
%
%
Finally, the multi-particle configuration interaction calculations with basis states provided by the above unfolding tight-binding method predict a very small fine-structure splitting of the ground state exciton both for bright and dark doublet. That highlights the possibility of the silicon G-center to be a very good emitter of entangled photons for quantum communication and computation applications.
\section{Acknowledgments}
\label{sec:acknowledgments}
The authors acknowledge funding from the European Innovation Council Pathfinder program under grant agreement No 101185617 (QCEED) and support by the project Quantum materials for applications in sustainable technologies, CZ.02.01.01/00/22 008/0004572. P.K. partly acknowledges funding by Institutional Subsidy for Long-Term Conceptual Development of a Research Organization granted to the Czech Metrology Institute by the Ministry of Industry and Trade of the Czech Republic.
%

%

%
%
\section*{Appendix I. }
\label{AppendixI}
%
\begin{figure*}[htbp]
    \includegraphics[width=150mm]{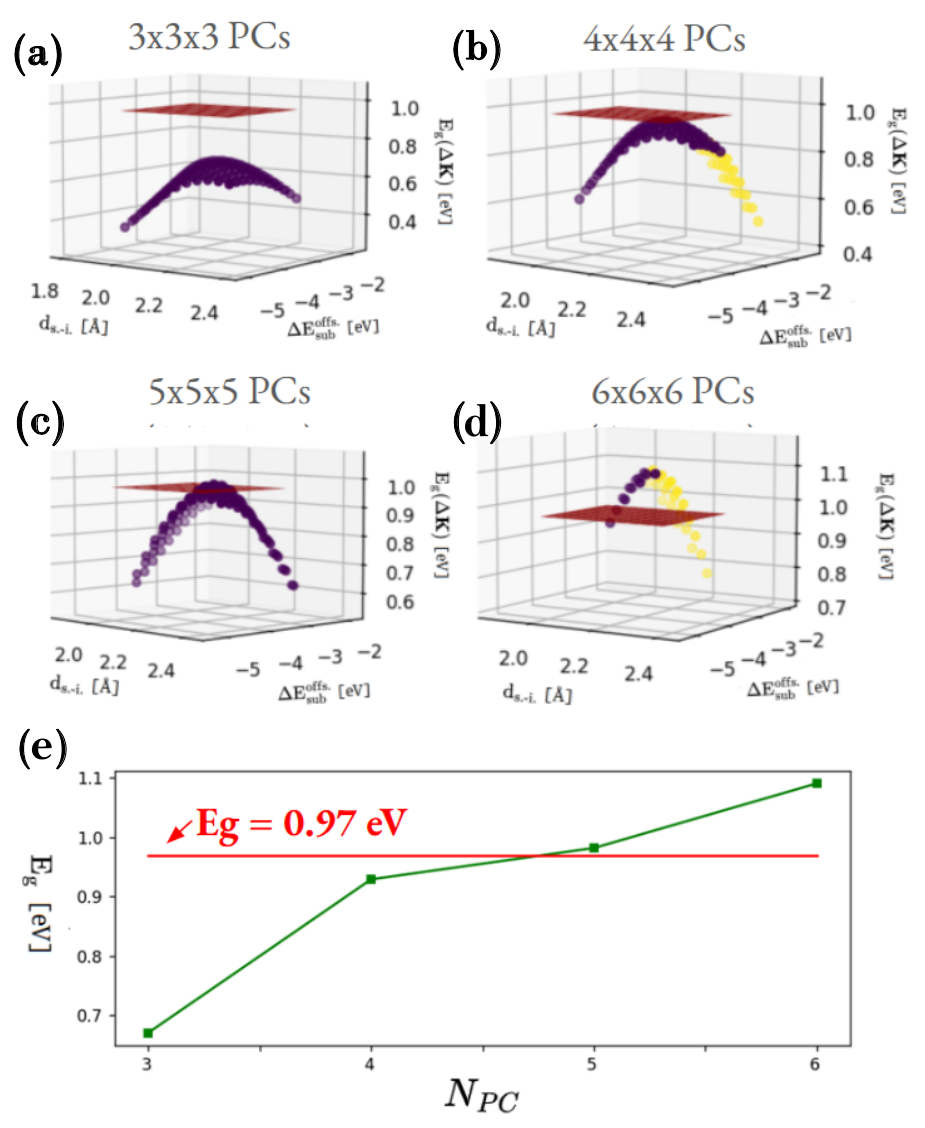}
	\caption{(a)-(d) Energy band gaps of the G-center ${E}_\textrm{g}(\Delta\mathbf{K})$ for i=0 and $N_{SC}=3,~4,~5,~6$ as a function the band offset of substitutional atom $\Delta {E}^{\rm offs.}_\textrm{sub}$ and the bond length ${d}_{\textrm{s.-i.}}$ where the ball colors mark a difference of reciprocal wavevectors during the transition (dark balls represents direct transition, yellow represents indirect transition). Red plane shows the energy of the G-center of $970$ meV. (e) Shows maximum band gaps as a function of $N_{SC}$ computed from data in panels (a)--(d) and red horizontal line represents the energy of the G-center.}
	\label{fig:SC_Eg}
\end{figure*}
%
We have calculated the evolution of band gap for different sizes of SCs ($N_{SC}=3,~4,~5,~6$) to analyze the evolution of the band gap of the G-center with a change of SC size and, consequently, with the density of the G-center in bulk silicon, see Fig.~\ref{fig:SC_Eg}. From Fig.~\ref{fig:SC_Eg}~(e), we see that the maximum band gap increases with SC size which may be understood as the G-center having a smaller impact on the band structure in larger supercells as one would generally expect. There is also a difference between results where in Fig.~\ref{fig:SC_Eg}~(b),~(d), where the transitions with indirect band gap (represented by yellow) appear due to asymmetry of SC (the G-center is not in the center of the structure). Clearly, the band gap of the G-center is influenced by size of the SC and its global symmetry.  

%

\end{document}